\documentclass[fleqn,usenatbib]{mnras}

\usepackage{newtxtext,newtxmath}

\usepackage[T1]{fontenc}

\DeclareRobustCommand{\VAN}[3]{#2}
\let\VANthebibliography\thebibliography
\def\thebibliography{\DeclareRobustCommand{\VAN}[3]{##3}\VANthebibliography}

\usepackage{graphicx}	
\usepackage{amsmath}	
\usepackage{xcolor}
\definecolor{RED}{HTML}{FF0000}

\title[Love numbers for interior retrievals]{Retrieving interior properties of hot Jupiters with Love numbers and atmospheric measurements}

\author[van Dijk et al.]{
E. A. van Dijk,$^{1}$\thanks{E-mail: evdijk@strw.leidenuniv.nl}
Y. Miguel$^{1, 2}$
\\
$^{1}$Leiden Observatory, Leiden University, Einsteinweg 55, 2333 CC, Leiden, the Netherlands\\
$^{2}$SRON Netherlands Institute for Space Research, Niels Bohrweg 4, 2333 CA, Leiden, the Netherlands\\
}

\date{Accepted XXX. Received YYY; in original form ZZZ}

\pubyear{2025}

\begin{document}
\label{firstpage}
\pagerange{\pageref{firstpage}--\pageref{lastpage}}
\maketitle

\begin{abstract}

Understanding exoplanet interiors is crucial for interpreting atmospheric observations and constraining their evolution and formation. However, due to limited observational constraints, interiors structures remain poorly understood. In this work, we investigate how new observational constraints, such as the Love number and atmospheric metallicity, improve our ability to characterize the interiors of hot Jupiters, planets for which Love number measurements are most feasible. We assess the precision required in Love number measurements to derive interior properties using both a simple two-layer homogeneous model and a more complex dilute core model. To account for observational uncertainties, we implement a retrieval framework. Our results show that accurately constraining core mass and bulk metallicity requires a high-precision Love number measurement, better than 40\% for a homogeneous model and 15\% for a dilute core model, along with an atmospheric metallicity measurement. We apply our retrieval framework to five planets with observed Love numbers, of which only WASP-19Ab has both an atmospheric metallicity constraint and a highly precise Love number measurement, with a precision of 12\%. For this flagship planet, both models confirm the presence of a core, although we cannot yet distinguish between a compact core or diluted core. With the homogeneous model, we find a core mass fraction of $0.21^{+0.05}_{-0.04}$, corresponding to $79^{+21}_{-18}$ $M_\mathrm{earth}$. Upcoming JWST observations are expected to provide high-precision Love number measurements and precise atmospheric data, offering new insights into the structure and composition of gas giant interiors. 

\end{abstract}






\section{Introduction}

The field of exoplanet science has rapidly evolved from detecting planets to characterizing their atmospheres and interiors in unprecedented detail. In particular, JWST's high-precision observations are transforming our understanding of exoplanet atmospheres. However, a planet’s atmospheric properties cannot be fully interpreted in isolation: they are shaped by its interior, which evolves over time and depends on its structure and composition. Unlike atmospheres, interiors cannot be directly observed and must be inferred by combining limited observational data with theoretical models, making them less well understood. Yet, studying planetary interiors is not only crucial for interpreting atmospheric properties but also for understanding planet formation and evolution. For instance, a planet’s bulk metallicity and distribution of heavy elements, provide key insights into its formation history, including mechanisms such as core accretion \citep[e.g.][]{Thorngren_2016, Hasegawa_2018, Hasegawa_2024}

Traditional interior models for exoplanets often assume a simple two-layer structure consisting of a dense heavy element core surrounded by a homogeneous hydrogen-helium-rich envelope \citep[e.g.][]{Miller_2011, Thorngren_2016, Muller_2024_2}. These studies constrain interior parameters by using observational data, such as mass, radius and age measurements, but the simplified interior models might become limiting as new observational constraints emerge. Recent data from JUNO and Cassini have demonstrated that Jupiter and Saturn have more complex interiors than these idealized models suggest, with evidence for diluted cores and inhomogeneous envelopes \citep[e.g.][]{Wahl_2017, Nettelmann_2021, Miguel_2022, Mankovich_2021}. These findings suggest that similar structures may exist in exoplanets and raise the question whether more advanced interior models are needed to interpret their observed properties. Some studies have incorporated additional constraints, such as atmospheric metallicity, into interior models \citep[e.g.][]{Bloot_2023}, demonstrating the potential of new observational constraints.

The Love number, which quantifies a planet's response to tidal forces, provides a direct constraint on the density distribution. Various studies have researched the applicability of the Love number in determining the interior structure of exoplanets, both for super-Earths \citep[e.g.][]{Kellermann_2018, Baumeister_2023} and gas giants \citep[e.g.][]{Kramm2011, Kramm2012ConstrainingHAT-P-13b, Padovan_2018, Wahl_2021}. In the simplest case of a two-layer model, the Love number is directly related to the core mass \citep{Kramm2011}. However, more complex three-layer models introduce degeneracies, reducing the direct correlation between Love number and core mass. A diluted core model falls between these two cases, and it remains unclear how precisely a Love number measurement can constrain such interiors. 

Due to strong tidal effects, Love number measurements are most feasible for hot Jupiters. Currently, the Love number has been measured for a sample of five exoplanets,  obtained through eccentricity measurements in a multiple planet system \citep[HAT-P-13b;][]{Batygin2009_firstkeccentricity, Buhler_2016, Hardy_2017}, transit timing variations caused by apsidal precession \citep[WASP-18Ab, WASP-19Ab; e.g.][]{Ragozzine_2009, love_number_wasp18b, Bernabo_2024} and direct light-curve deformation \citep[WASP-12b, WASP-103b; e.g.][]{Hellard_2019, Barros_2022, Akinsanmi-2024-2}. Additionally, \citet{Hellard_2020} found a tentative detection for WASP-121b, but due to the large errorbars, we do not include this planet in our work. Given these ongoing and upcoming Love number measurements, it is important to estimate the precision required for Love numbers to meaningfully improve interior constraints. 

Previous studies have largely focused on theoretical modelling without fully incorporating observational uncertainties, such as those associated with mass, radius, inflation and the Love number. In this paper, we study the role of new observational constraints, specifically atmospheric metallicities and Love numbers, in retrieving interior properties of hot Jupiters, while accounting for observational uncertainties. We compare constraints obtained from a simple two-layer homogeneous model with those from a more complex diluted core model to evaluate the effectiveness of Love numbers in modelling more complex interior structures. Additionally, we assess the minimum Love number precision required to improve interior constraints, providing guidance for future observations. Lastly, we apply our retrieval framework to exoplanets with observed Love numbers to determine their interior structures.
\section{Methods} 

\subsection{Interior model} 

The interior structure is modelled with CEPAM \citep{CEPAM_Guillot}. This code has been used to model both planets in our solar system \citep[e.g.][]{Miguel_2022} and extrasolar planets \citep[e.g.][]{Valencia_2013, Bloot_2023}. It numerically solves the interior structure equations, i.e. the equation of hydrostatic equilibrium, energy transport, mass conservation and energy conservation. 

\begin{figure*}
    \centering
    \includegraphics[width=0.6\textwidth]{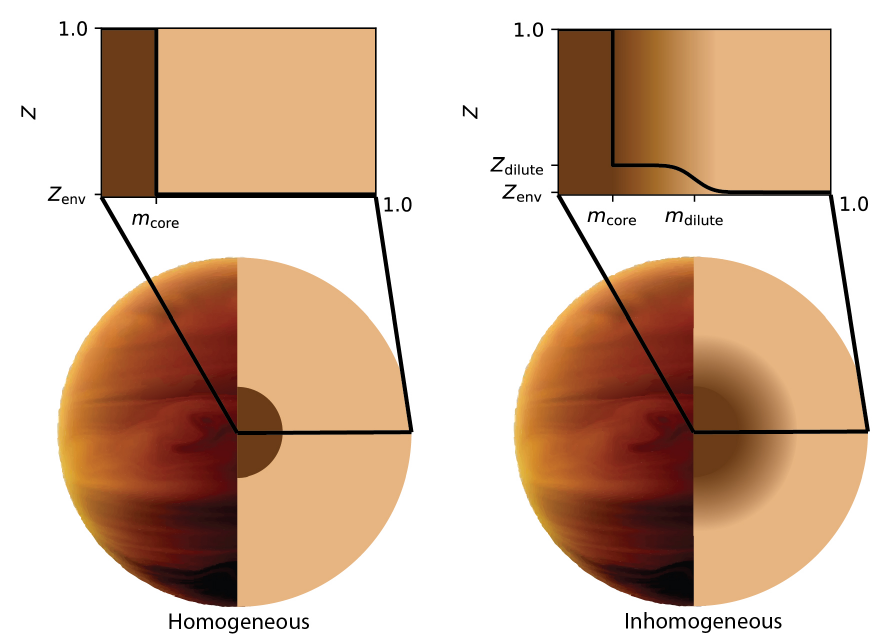}
    \caption{The two different interior structure models used in this work. Both models have an isothermal rocky core, with a heavy element fraction of 1. In the homogeneous model (left) there is a sharp transition at $m_\mathrm{core}$ to the envelope, consisting out of hydrogen (X), helium (Y) and heavy elements ($Z_\mathrm{env}$). In the inhomogeneous model (right) we add a dilute core region, where the heavy element fraction reduces gradually from $Z_\mathrm{dilute}$ to $Z_\mathrm{env}$. The mass coordinate $m_\mathrm{dilute}$ controls the extent of the dilute core.}
    \label{fig:schematic_hom_inhom}
\end{figure*}

CEPAM can model a homogeneous and inhomogeneous interior structure \citep{Miguel2016, Miguel_2022} (see Figure \ref{fig:schematic_hom_inhom}). The homogeneous interior structure consists of a core of heavy elements and a helium- and hydrogen-rich homogeneous envelope with some heavy elements, represented by water. This model is characterized by the parameters $m_\text{core}$, the core mass fraction, $Z_\text{env}$, the metal mass fraction in the envelope, and X and Y, the hydrogen and helium mass fractions. The inhomogeneous model is similar but adds a dilute core region between the homogeneous envelope and the heavy element core with a heavy element fraction that reduces gradually from the maximum heavy element fraction of the dilute core $Z_\mathrm{dilute}$ to the heavy element fraction in the envelope $Z_\mathrm{env}$. The gradient is the steepest at the mass coordinate $m_\mathrm{dilute}$. This results in a heavy element fraction at mass coordinate $m$ of

\begin{equation}
    Z(m) = Z_{\text{atm}} + \frac{Z_\text{dilute} - Z_{\text{atm}}}{2} \left[1 - \text{erf}\left(\frac{m - m_\text{dilute}}{\partial m}\right)\right].
\end{equation}

In this equation, $\partial m$ defines the width of the heavy element gradient, which we, similarly to \citet{Bloot_2023}, set to $0.075$.

The default equations of state used in this work are \citet{EoS_hydrogen1} and \citet{Miguel2016} for hydrogen (MH13-H),  \citet{EoS_helium} for helium (SCH95-He) and \citet{Mazevet_2019_2} for water in the envelope. For the rocky core we use the analytical approximation by \citet{Hubbard_1989}. We expect that differences that may arise by using a different hydrogen and helium equation of state will be smaller than differences due to observational uncertainties for the mass range of planets that we are modelling \citep{Howard_2024_2}.

\subsubsection{Atmospheric boundary conditions}
The boundary of the interior is defined by the atmosphere. This is modelled with a non-grey opacity model \citep{Parmentier_2014, Parmentier_2015}. The composition of this model atmosphere is set to solar and includes TiO and VO. However, in the case our envelope metallicity is super- or sub-solar, we use \citet{Valencia_2013}'s approach to interpolate the Rosseland opacities to be consistent with a higher or lower metallicity atmosphere. The optical depth limit of the atmosphere is set to $10^2$. Varying this parameter between 1 and $10^3$ results in differences in the radius and Love number of less than 1\% and 7\%. Other parameters are set to the default values specified by \citet{Parmentier_2014}. 

\subsubsection{Additional heating} \label{sec:heating}
It has been long known that gas giants with high equilibrium temperatures ($T_\mathrm{eq}$>1000 K), commonly known as hot Jupiters, have higher radii than standard evolution models predict \citep{Guillot_2002, Demory_2011, Miller_2011}. To date, all planets with measured Love numbers are hot Jupiters. Therefore, this inflation effect must be taken into account. Various mechanisms have been proposed to explain the inflation effect. Those mechanisms can be divided in mechanisms related to tidal heating, such as tidal dissipation \citep[e.g.][]{Bodenheimer-2001} and thermal tides \citep{Arras-2010}, and mechanisms that convert a fraction of the incident stellar flux to intrinsic heat, such as atmospheric circulation \citep{Guillot_2002, Showman-2002}, ohmic dissipation \citep{Batygin-2010, Batygin-2011, Perna-2010, Huang-2012, Wu-2013, Ginzburg-2016}, advection of potential temperature \citep{Tremblin-2017, Sainsbury-Martinez-2019} and the mechanical greenhouse \citep{Youdin-2010}. In this work, we do not assume a specific mechanism responsible for the inflation. Instead, we incorporate it into our model by introducing a free heating parameter that scales with stellar irradiation $L_\mathrm{irr}$. The fraction of the stellar irradiation that is converted to extra heat, which we denote as $\gamma$, can range between $10^{-5}$ and 0.1 (similar to \citet{Komacek_2017}). We assume that this additional heat is uniformly distributed throughout the planet’s interior and contributes to its total luminosity, which consists of both this extra heat and the internal luminosity the planet would have at its evolutionary stage if it were not inflated. We denote this intrinsic, non-inflated luminosity as $L_\mathrm{grav}$. Then the total luminosity of the planet is given by:

\begin{equation}
    L_\mathrm{int} = L_\mathrm{grav} + \gamma L_\mathrm{irr}.
\end{equation}

\subsubsection{Love number calculation} \label{sec:methods_lovenumber}

The Love function and Love number characterize how a planet's gravitational potential responds to an external tidal perturbation, such as that induced by a nearby star \citep{love_1911}. The Love function, $K_{n}(r)$, is defined as
\begin{equation}
    V_{n, \mathrm{ind}}(r) = K_{n}(r) W_{n}(r),
\end{equation}
where $V_{n, \mathrm{ind}}(r)$ is the induced gravitational potential and $W_{n}(r)$ is the external tidal potential. The Love number represents the value of the Love function evaluated at the planetary surface, expressed as $K_{n} (R_p) = k_{n}$. We focus on the first-order tidal response, denoted as $k_{2}$, which is directly linked to planet's internal mass distribution. A more uniform mass distribution results in a larger tidal response and, consequently, a higher Love number. The theoretical maximum is 1.5, corresponding to a fully uniform density profile. In contrast, a planet with a more centrally concentrated mass experiences a weaker tidal response, leading to a lower Love number, with a minimum value of 0. 

Assuming hydrostatic equilibrium and neglecting higher-order non-linear effects, the first-order Love number can be determined solely from the density profile. The Love number, which corresponds to twice the apsidal motion constant described in \citet{Sterne}, is given by 

\begin{equation}
    k_{2} = \frac{3 - \eta_2(R_{p})}{2 + \eta_2(R_{p})},
\end{equation}

where $\eta_2(R_p)$ is obtained by solving the following ordinary differential equation for $\eta_2(r)$, integrated radially outward from the planetary centre with the initial condition $\eta_2(0) = 0$. 

\begin{equation}
    r \frac{\mathrm{d}\eta_2(r)}{\mathrm{d}r}  + \eta_2(r)^2 - \eta_2(r) - 6 + \frac{6\rho(r)}{\rho_m(r)} (\eta_2(r)+1) = 0
\end{equation}

Here, $\rho (r)$ is the local density at radius $r$, as computed from the planetary interior model, and $\rho_m$ is the mean density enclosed within radius $r$.

\subsection{Retrieval framework}

We use a Bayesian framework to fit our data to the static interior model, for which we use PyMultinest \citep{pymultinest, multinest}. This is a Nested Sampling Monte Carlo algorithm and improves on traditional Monte Carlo Markov Chain (MCMC) methods by being more suited to handle degeneracies in the resulting posterior distributions. We use 1000 live points and a sampling efficiency of 0.1.

We assume normally distributed uncertainties for both the radius and Love number of the planet and calculate the log-likelihood using the calculated radius and Love number.

\begin{equation}
    \ln{\mathcal{L}(\theta)} = - \frac{1}{2} \sum_{X\in\{R_p, k_{2}\}}\left[\frac{(X_{\text{observed}} - X_{\text{model}})^2}{\sigma_X^2} + \ln{(2\pi\sigma_X^2)} \right]
\end{equation}

The variable $X$ denotes the data, consisting of the radius $R_p$ and the Love number $k_{2}$. The variables in $X_\text{model}$ are calculated using the model with parameters $\theta$. $X_\text{observed}$ are the observed radius and Love number of the planet and $\sigma_X$ indicates the uncertainty in the respective variables.

\subsection{Priors} 

The input parameters consist of the varying parameters $\theta$, which for the homogeneous and inhomogeneous model are set to

\begin{equation}
\begin{cases}
\theta_\mathrm{homogeneous} = \{M_p, m_\mathrm{core}, Z_\mathrm{env}, T_\mathrm{eq}, L_\mathrm{grav}, \gamma\} \\ \theta_\mathrm{inhomogeneous} = \{M_p, m_\mathrm{core}, Z_\mathrm{env}, m_\mathrm{dilute}, Z_\mathrm{dilute}, T_\mathrm{eq}, L_\mathrm{grav}, \gamma \}
\end{cases}
\end{equation}

and the data $X = \{R_p, k_{2}\}$. The amount of hydrogen and helium in the envelope is set by the metal mass fraction in the envelope by assuming a proto-solar ratio between hydrogen and helium \citep{Lodders_2021}. In this way we reduce the compositional freedom in the envelope to only one parameter: Z$_\mathrm{env}$. 

We impose a Gaussian prior on a parameter if this parameter is observationally constrained, where the Gaussian is centred around the measured value and the width reflects the uncertainty of the measurement. This is the case for the planet mass $M_\mathrm{p}$ and equilibrium temperature $T_\mathrm{eq}$. The prior on the envelope metallicity $Z_\mathrm{env}$ can be either observationally constrained or physically constrained. When we have an observational estimate of the atmospheric metallicity we impose a Gaussian prior and assume that the atmospheric metallicity reflects the metallicity in the envelope. Observationally determined metallicity constraints are often given in number densities relative to solar instead of mass fractions, the unit used in our interior models. In Appendix \ref{ap:metallicity}, we show how we convert the observational constraints to mass fractions. When the atmospheric metallicity is unknown we use a physical constraint: a uniform prior between 0 and 1. Similarly, we impose uniform physically constrained priors for the core mass fraction $m_\mathrm{core}$ and the dilute core parameters $m_\mathrm{dilute}$ and $Z_\mathrm{dilute}$. All priors are summarized in Table \ref{tab:priors}. 

\begin{table}
    \centering
    \caption{The priors for the free parameters in the retrievals. The Gaussian priors are based on observational constraints. The other parameters are physically constrained with a uniform or log-uniform prior. }
    \begin{tabular}{l|c}
    \hline
     Parameter         &  Prior \\ \hline
     $M_\mathrm{p}$             &  $\mathcal{N}(\mu_{M_p}, \sigma_{M_p}$) \\
     $T_\mathrm{eq}$   &  $\mathcal{N}(\mu_{T_\mathrm{eq}}, \sigma_{T_\mathrm{eq}})$ \\
      $Z_\mathrm{env}$ (observed/physical)  &  $\mathcal{N}(\mu_{\mathrm{[M/H]}}, \sigma_{\mathrm{[M/H]}})$ / $\mathcal{U}(0, 1)$      \\
     $L_\mathrm{grav}$ [$L_\mathrm{J}]$ &  $\mathcal{L}\mathcal{U}(0, l_\mathrm{agemin})$ / $\mathcal{U}$(1, $L_\mathrm{agemin}$)\\
     $\gamma$ & $\mathcal{L}\mathcal{U}(-5, -1)$ / $\mathcal{U}$($10^{-5}, 0.1)$\\
    $m_\mathrm{core}$ &  $\mathcal{U}(0, 1)$ \\ \hline
     $m_\mathrm{dilute}$ & $\mathcal{U}$($m_\mathrm{core}$, 1) \\
     $Z_\mathrm{dilute}$ & $\mathcal{U}$($Z_\mathrm{env}$, 1) \\ \hline
    \end{tabular}

    \label{tab:priors}
\end{table}

To avoid a nonphysical solution, we impose some additional constraints, similar to \citet{Bloot_2023}. First of all, we require the sum of the helium fraction $Y$ and the envelope metal fraction $Z_\mathrm{env}$ to be smaller or equal to $0.75$. Using this, we avoid too large values of $Y$ and $Z_\mathrm{env}$ and ensure convergence. Secondly, we require $m_\mathrm{dilute}$ to be larger than $m_\mathrm{core}$ and $Z_\mathrm{dilute}$ to be larger than $Z_\mathrm{env}$.

The prior on the internal luminosity requires some more explanation. As explained in Section \ref{sec:heating}, the internal luminosity consists of two free parameters: the extra heating efficiency $\gamma$ and the luminosity due to the planets contraction and evolution $L_\mathrm{grav}$. We use \citet{Bloot_2023}'s approach to constrain $L_\mathrm{grav}$ from evolutionary models and the stellar age. The lower bound on the age gives us an upper luminosity limit. The lower limit we set to 1 $L_\mathrm{J}$. Similarly to \citet{Komacek_2017}, we let $\gamma$ vary between $10^{-5}$ and 0.1.

We use both a linear-uniform and log-uniform prior for both $L_\mathrm{grav}$ and $\gamma$. There is no physical reason to choose one or the other, but the choice of prior influences the results we find for some planets. In Section \ref{sec:test_planet} and \ref{sec:real_planets} we only show the results that use the log-uniform prior. The differences between the priors are discussed in Appendix \ref{ap:internal_luminosity}.
\section{Test planet} \label{sec:test_planet}

\begin{table}
    \caption{The properties of the test planet. The reported errors on the mass, radius, equilibrium temperature, atmospheric metallicity and age are consistent with current observations.}
    \begin{tabular}{l|c}
    \hline
     Observational constraints    &  Test value\\ \hline
     $M_\mathrm{p}$ ($M_\mathrm{J}$)    & $1.00\pm0.06$ \\
     $R_\mathrm{p}$ ($R_\mathrm{J}$) & $1.47\pm0.05$ \\
     $T_\mathrm{eq}$ (K) & $2000\pm35$ \\
     Stellar age $t_*$ (Gyr) & $4.0\pm0.5$ \\
     $Z_\mathrm{env}$ ([M/H], dex) & $-0.0338\pm0.305$ \\
     $k_{2}$ & $0.32\pm0.06$ \\
     $F_\mathrm{inc}$ ($10^9$ erg s$^{-1}$ cm$^{-2}$) & 3.629 \\
    \hline
    Interior properties & \\ \hline
    $Z_\mathrm{env}$ & 0.016 \\
    $m_\mathrm{core}$ & 0.006 \\
    $m_\mathrm{dilute}$ & 0.26 \\
    $Z_\mathrm{dilute}$ & 0.18 \\
    $L_\mathrm{grav}$ (L$_\mathrm{J}$) & 1.96 \\
    $\gamma$ & 0.001 \\
    $Z_\mathrm{planet}$ & 0.0636 \\ \hline
    \end{tabular}

    \label{tab:test_planet}
\end{table}

\begin{figure*}
    \centering
    \includegraphics[width=\textwidth]{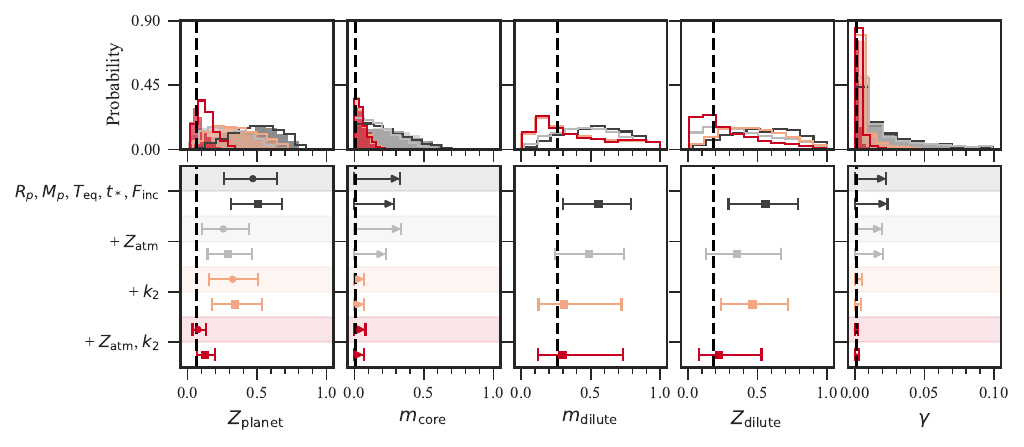}
    \caption{Posterior distributions (upper panels) and summary statistics (lower panels) of the bulk metal fraction ($Z_{\mathrm{planet}}$), core mass fraction ($m_{\mathrm{core}}$), dilute core parameters ($m_{\mathrm{dilute}}$, $Z_\mathrm{dilute}$) and heating efficiency ($\gamma$) for the test planet with properties in Table \ref{tab:test_planet}. The summary statistics are either represented by a median with the 16th to 84th percentiles (for $Z_\mathrm{planet}$, $m_\mathrm{dilute}$ and $Z_\mathrm{dilute}$) or a 2-sigma (68th percentile) upper limit (for $m_\mathrm{core}$ and $\gamma$). We show, for both the homogeneous (shaded) and inhomogeneous (not shaded) model, retrieval results where we include constraints on the mass, radius, equilibrium temperature and stellar age (black) and retrieval results where we add an additional constraint on the atmospheric metal fraction (grey), the Love number (orange) or both (red). The dashed black lines indicate the true interior properties of the planet from Table \ref{tab:test_planet}.}
    \label{fig:interior_params_posteriors_test_planet}
\end{figure*}

To test the model and examine the impact of including the observables $k_{2}$ and $Z_\mathrm{atm}$ in the retrieval process, we create a test planet. The properties of this planet are summarized in Table \ref{tab:test_planet}. For this planet, we assume an inhomogeneous interior structure resembling (one of the possible configurations of) Jupiter's interior \citep{Miguel_2022}. However, we introduce extra heat to the interior to artificially inflate the planet and we raise its equilibrium temperature, making it more analogous to the hot planets with measured Love numbers. The assumed uncertainties for mass, radius, equilibrium temperature, atmospheric metallicity and stellar age are consistent with current exoplanetary observations (see examples in Table \ref{tab:dataset}). 

When comparing the homogeneous and inhomogeneous model we need to take into account that two planets with the same bulk metallicity but a different interior structure model will result in a slightly different radius and Love number. For this test planet, a homogeneous model with the metals of the original dilute core moved to the compact core will result in a radius and Love number that are respectively 1\% and 4\% smaller than for the original inhomogeneous model. In a perfect scenario, this will cause the retrieved core mass of the homogeneous model to be slightly higher and the retrieved bulk metallicity to be slightly lower.

\subsection{Importance of measurements for $Z_\mathrm{atm}$ and $k_{2}$} \label{sec:testplanet_adding_z_k}

We investigate how crucial it is to include the atmospheric metallicity and the Love number in determining the interior structure of a hot Jupiter. To do this, we conduct retrieval analyses on a test planet, both with and without these measurements, using homogeneous and inhomogeneous models. While all models can reproduce the observed properties such as planetary mass, radius, equilibrium temperature and Love number, we observe significant differences in the posterior distributions of the retrieved interior parameters. These posteriors (upper panels), along with summary statistics (lower panels), are shown in Figure \ref{fig:interior_params_posteriors_test_planet}. The true parameter values given to the test planet are indicated by black dashed lines for comparison. The results retrieved with the homogeneous/inhomogeneous model are shown as filled/clear histograms in the upper panel and with a shaded/unshaded background in the lower panel.

\subsubsection{Bulk metallicity}
An accurate estimation of bulk metallicity, shown in the left panel in Figure \ref{fig:interior_params_posteriors_test_planet}, requires measurements of both atmospheric metallicity and Love number. Only when both of these measurements are incorporated (shown in red in Figure \ref{fig:interior_params_posteriors_test_planet}), we can retrieve the bulk metallicity with the homogeneous model within one sigma from the real value. The precision of the bulk metallicity also significantly improves from 39\% to 10\% of the total parameter space. Compared to the homogeneous model, the inhomogeneous model provides a less precise and accurate constraint, due to the larger number of free parameters. When incorporating all observational constraints, the retrieved value for the inhomogeneous model is slightly more than one sigma away from the actual value but remains within two sigma. The dispersion in the distribution reduces from 37\% to 12\% of the parameter space in this case. As we will demonstrate in Section \ref{sec:uncertainty_love_number}, improving the precision of the Love number can further improve the accuracy of this constraint, bringing it within one sigma agreement with the actual value. 

\subsubsection{Core mass fraction}
For a homogeneous model the Love number is expected to directly relate to the core mass \citep{Kramm2011}, and our results, shown in the second panel of Figure \ref{fig:interior_params_posteriors_test_planet}, align with this expectation. Including a measurement of the Love number (shown in orange and red in Figure \ref{fig:interior_params_posteriors_test_planet}) significantly improves the constraints on the core mass, reducing the 2-sigma upper limit from 105 $M_\oplus$ to 23 $M_\oplus$, which is equal to constraining the core mass from 33\% to 7\% of the complete parameter space. For the inhomogeneous model, we observe a similar effect on the core mass. With the Love number measurement, the 2-sigma upper limit on the core mass decreases from 88.6 $M_\oplus$ to 21.7 $M_\oplus$, which is equal to a reduction from 28\% to 7\% of the complete parameter space.

The inhomogeneous model provides slightly better constraints on the core mass fraction compared to the homogeneous model. In general, we primarily obtain an upper limit on the core mass fraction rather than a precise value for this test planet. In the inhomogeneous model, the metals are distributed across both the compact and dilute core, whereas in the homogeneous model, they are confined exclusively to the compact core. As a result, the inhomogeneous model consistently predicts a lower core mass fraction, resulting in a lower upper limit.

\subsubsection{Dilute core parameters}
The dilute core of the inhomogeneous model is characterised by two parameters - its extent and its metal content - depicted in the third and fourth panels of Figure \ref{fig:interior_params_posteriors_test_planet}. The extent of the dilute core, defined by $m_\mathrm{dilute}$, is more accurately estimated when the Love number is included, with the true value falling within one sigma of the retrieved parameters, despite a broad dispersion. For the maximum metallicity of the dilute core, both the Love number and atmospheric metallicity constraints improve parameter estimation. When an atmospheric metallicity measurement is included, the true value falls within one sigma of the retrieved values. Adding a Love number constraint further improves the precision of the retrieved parameter. 

\subsubsection{Heating efficiency}
With the Love number and atmospheric metallicity measurement, we can also constrain the heating efficiency, shown in the right panel of Figure \ref{fig:interior_params_posteriors_test_planet}. With the constraint on the Love number and atmospheric metallicity, we retrieve a 2-sigma upper limit of 0.002 for the heating efficiency fraction for the homogeneous model and an upper limit of 0.003 for the inhomogeneous model, compared to 0.023 (homogeneous) and 0.024 (inhomogeneous), when these measurements are not included.

\subsection{Uncertainty in the Love number} \label{sec:uncertainty_love_number}

\begin{figure}
    \centering
    \includegraphics[width=\columnwidth]{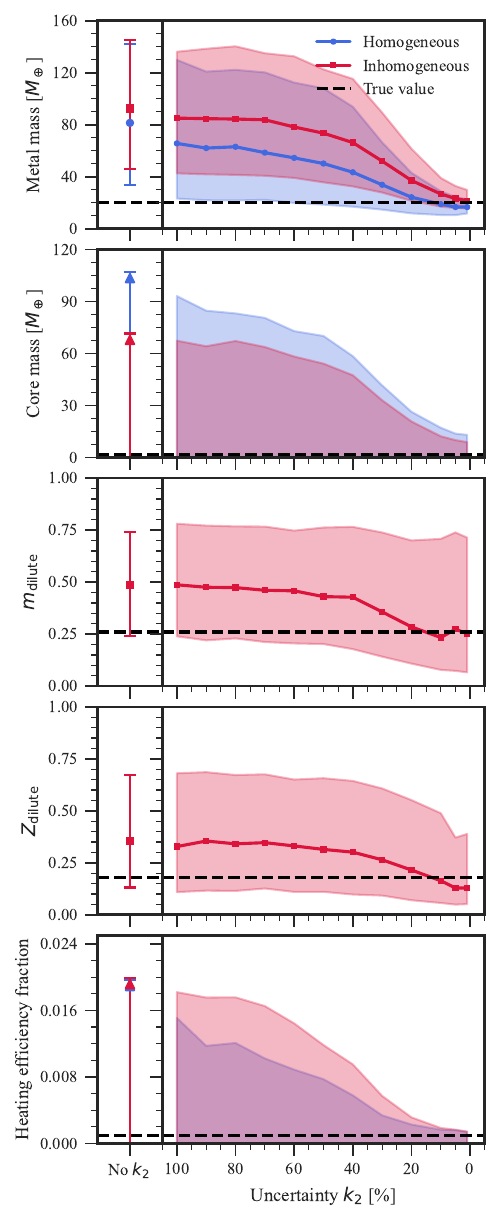}
    \caption{The predicted posteriors of the interior parameters resulting using different uncertainties in the Love number as input for the test planet. The posteriors are represented by the median and 1-sigma confidence intervals for the metal mass and dilute core parameters and by 2-sigma upper limits for the core mass and heating efficiency. The dashed black lines indicate the true values. Results for the homogeneous (blue) and inhomogeneous model (red) are shown.}
    \label{fig:test_planet_errors_k}
\end{figure}

\begin{figure}
    \centering
    \includegraphics[width=0.49\columnwidth]{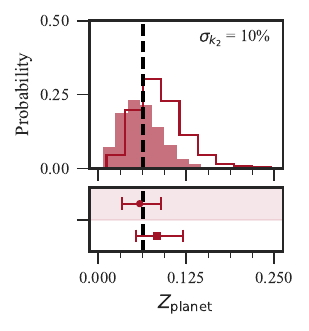}
    \includegraphics[width=0.49\columnwidth]{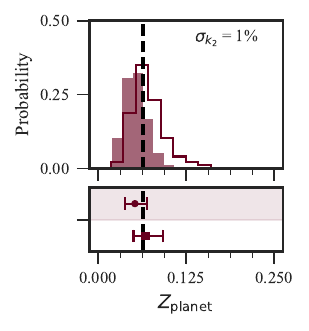}
    \caption{The posterior distributions (upper panels) and corresponding confidence intervals (lower panels) of the bulk metallicity resulting from retrievals with Love number uncertainties of 1 and 10\%. Results of the homogeneous (shaded) and inhomogeneous (not shaded) model are compared. The dashed black lines indicate the true values.}
    \label{fig:test_planet_small_errors_k}
\end{figure}

In the previous section, the uncertainty in the Love number was fixed to 20\%. However, to investigate how the precision of the Love number measurements impacts interior modelling, we varied its uncertainty from 1\% to 100\%. For these runs, we included the measurement of atmospheric metallicity. The resulting posterior distributions, represented by their median and 16th to 84th percentiles for the metal mass and dilute core parameters and by 2-sigma upper limits for the core mass and heating efficiency, are shown in Figure \ref{fig:test_planet_errors_k}. 

As expected, for both for the homogeneous and inhomogeneous models, the uncertainties in the core mass and the bulk metal mass, decrease as the precision of the Love number measurement improves. A high uncertainty of 100\% in the Love number provides no meaningful improvement in the inferred interior parameters compared to having no Love number measurement at all. Reducing the uncertainty to 50\% yields modest improvements, but the most significant benefits are observed when the uncertainty drops below 40\%. A similar trend is seen for the heating efficiency.

In Section \ref{sec:testplanet_adding_z_k}, we could not retrieve the bulk metallicity within one sigma of the true value with the inhomogeneous model. In the first panel of Figure \ref{fig:test_planet_errors_k} it is shown that we can retrieve the true bulk metallicity within one sigma if the Love number uncertainty is further reduced to 15\% or lower. 

Interestingly, the dilute core parameter $m_\mathrm{dilute}$ remains largely unaffected until the Love number uncertainty is reduced to 30\%. While the median of the posterior becomes more accurate, the posterior still exhibits a wide spread, with predictions between 0 and 1 remaining consistent with the observations. Further reducing the uncertainty below 20\% does not result in significant changes. For the dilute core parameter $Z_\mathrm{dilute}$, however, the confidence interval continues to shrink as the Love number uncertainty decreases until an uncertainty of 10\%.

We reiterate that both the homogeneous and inhomogeneous model can explain the observations, even as the error on the Love number decreases to 1\%. The retrieved interior parameters remain consistent with the true values within their uncertainties. However, as discussed in Section \ref{sec:test_planet}, there are discrepancies between the radius and Love number for homogeneous and inhomogeneous interiors with the same bulk metallicity, differing by approximately 1\% and 4\%, respectively. Our results do show some evidence of this discrepancy.

Up to a Love number uncertainty of 10\%, the homogeneous model provides a more accurate and precise estimate of the bulk metallicity compared to the inhomogeneous model, with its median value closer to the true value and a smaller associated uncertainty. However, this trend shifts as the uncertainty decreases even further. This effect is illustrated more clearly in Figure \ref{fig:test_planet_small_errors_k}, where we focus on runs with Love number uncertainties of 10\% and 1\%. At the smallest uncertainty, both models retrieve metal masses that agree with the true value within their respective uncertainties. However, while the inhomogeneous model's peak aligns precisely with the true value, the homogeneous model's peak slightly underestimates bulk metallicity as expected, since the test planet was designed with an inhomogeneous interior.

\section{Sample of exoplanets} \label{sec:real_planets}

In this section, we present the inferred interior properties of a sample of exoplanets with measured Love numbers. We represent each parameter by the median and 16th to 84th percentile of the retrieved posterior distribution. However, if the posterior distribution is skewed towards the lower prior boundary, we instead provide an upper limit where appropriate. The observational properties of the planets are summarized in Table \ref{tab:dataset}. We take masses, radii, equilibrium temperatures, atmospheric metallicities, stellar ages and Love numbers directly from the previous studies referenced in Table \ref{tab:dataset} and calculate the incident flux, $F_\mathrm{inc}$, from the stellar temperature, stellar radius and semi-major axis for each planet.

\begin{table*}    
    \caption{Properties of the exoplanets in our sample. References: 1:\citet{Buhler_2016}, \citet{Turner_2016}, \citet{Bonomo_2017}, 2: \citet{chakrabarty-2019}, \citet{Edwards-2023}, \citet{Akinsanmi-2024-2}, 3: \citet{Cortes_Zuleta_2020}, \citet{Goyal_2021}, \citet{Coulombe_2023}, \citet{love_number_wasp18b}, 4: \citet{Cortes_Zuleta_2020}, \citet{Edwards-2023}, \citet{Bernabo_2024} 5: \citet{wasp103b_data}, \citet{Goyal_2021}, \citet{Edwards-2023}, \citet{Barros_2022}.}
    \centering 
    \begin{tabular}{l|rrrrrrrr}
    \hline 
    Name            & Mass                      & Radius                      & $T_{\text{eq}}$ & [M/H]            & Stellar age             & $k_{2}$      & $F_\mathrm{inc}$         \\
                     & $M_J$                    & $R_J$                       & K               &   dex          & Gyr                     &             & $10^{9}$ erg s$^{-1}$ cm$^{-2}$           \\
    \hline
    HAT-P-13b$^{1}$ & $0.899^{+0.030}_{-0.029}$  & $1.487\pm0.041$            &  $1740\pm27$    & -                            &  $5^{+2.50}_{-0.70}$    & $0.31^{+0.08}_{-0.05}$ & 1.672\\
    
    WASP-12b$^{2}$  & $1.465\pm0.079$            & $1.937\pm0.056$            & $2592.6\pm57.2$ & $-0.07^{+0.63}_{-0.49}$ & $2.00^{+0.70}_{-2.00}$  & $0.55^{+0.45}_{-0.49}$ & 10.001\\
    
    WASP-18Ab$^{3}$  & $10.20\pm0.35$             & $1.240\pm0.079$           & $2416\pm58$     & {$0.013\pm 0.307$}$^*$        &  $1.57^{+1.40}_{-0.94}$ & $0.62^{+0.55}_{-0.19}$ & 8.900\\

    WASP-19Ab$^{4}$  & $1.154^{+0.078}_{-0.080}$  & $1.415^{+0.044}_{-0.048}$ & $2113\pm29$     & $0.95^{+0.57}_{-1.11}$  & $6.40^{+4.10}_{-3.50}$  & $0.20^{+0.02}_{-0.03}$ & 4.524\\
    WASP-103b$^{5}$ & $1.49\pm0.088$             & $1.528^{+0.073}_{-0.047}$   & $2513\pm49$     & $-0.3^{+0.48}_{-0.42}$          & $4.0\pm1.0 $            & $0.59^{+0.45}_{-0.53}$ & 6.456\\
    \hline
    
    \end{tabular}
    \footnotesize{\\ $^*$ WASP-18Ab's metallicity is re-calculated to a log-scale from the value in \citet{Coulombe_2023} of M/H = $1.03^{+1.11}_{-0.51}$ times solar.}
    \label{tab:dataset}
\end{table*}

\subsection{Reproducing the measurements} \label{sec:reproducing_measurements}

\begin{figure*}
    \centering
    \includegraphics[width=0.32\textwidth]{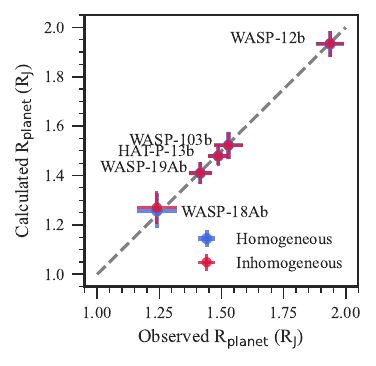}
    \includegraphics[width=0.32\textwidth]{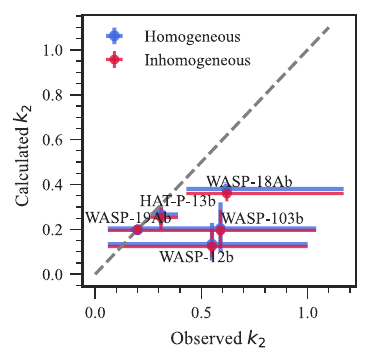}
    \includegraphics[width=0.34\textwidth]{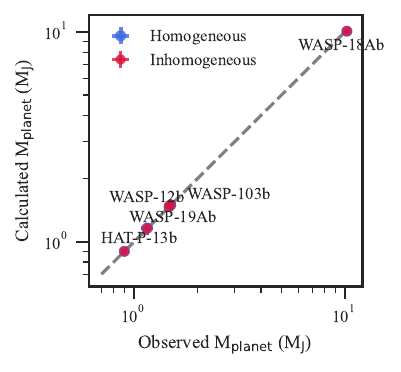}
    \caption{The planetary radius (left), Love number $k_{22}$ (middle) and planetary mass (right) that the model predicts and the observed values for the five planets. The results for both the homogeneous and inhomogeneous model are shown.}
    \label{fig:accuracy_fits}
\end{figure*}

To assess the accuracy of our fits, we compare the masses, radii and Love numbers that our framework predicts with the observed data in Figure \ref{fig:accuracy_fits}. The predicted radii and masses show good agreement with the observed values, remaining within the measurement uncertainties for all planets for both the homogeneous and inhomogeneous model. However, the predicted Love numbers, shown in the middle panel, deviate more from the observed values. For planets with the most precise Love number measurements, such as WASP-19Ab and HAT-P-13b, the predictions align well with observations. The retrieved Love numbers of WASP-103b and WASP-12b also agree within one sigma, because the observational uncertainties are relatively high for these planets. For WASP-18Ab, the predicted Love number falls slightly outside of the 1-sigma range but remains within two sigma. 

In general, our retrieved Love numbers are either at the lower end of what observations measure or lower than what observations suggest. This discrepancy was earlier also noticed by \citet{Wahl_2021}, who modelled the tidal response of various hot Jupiters (including HAT-P-13b, WASP-12b, WASP-18Ab and WASP-103b) with the more accurate concentric MacLaurin spheroid method. The theoretical ranges they calculate for HAT-P-13b, WASP-12b, WASP-18Ab and WASP-103b agree with the values we obtain, indicating that the discrepancy is not due to the non-linear effects that we are not taking into account. The difference between the theoretical and observational Love numbers could indicate either a systematic overestimation of the Love number in the observational methods or a missing physical mechanism that would enhance the Love number. Such a mechanism could be either fast rotation, resulting in non-tidally locked hot Jupiters, or dynamical contributions to the Love number. For instance, the dynamical contributions for Jupiter result in a 4\% lower Love number than the purely hydrostatic value \citep[e.g.][]{Idini-2021}. However, for these exoplanets, the contribution would need to enhance the Love number and should be significantly larger. It is worth noting that we find the largest deviations for planets with large uncertainties in the observations. Future high-precision observations will be able to further confirm or contradict these high Love number observations.

\subsection{Interior structure parameters}

\begin{figure*}
    \centering
    \includegraphics[width=\columnwidth]{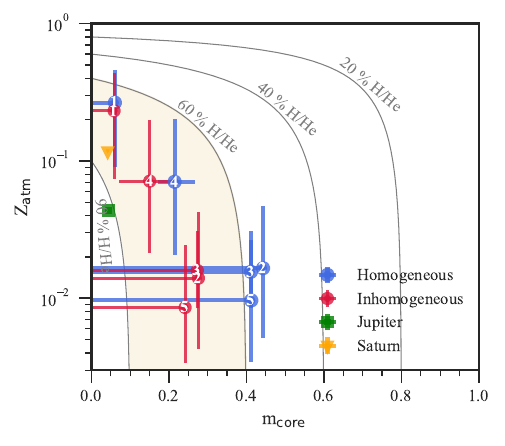}
    \includegraphics[width=\columnwidth]{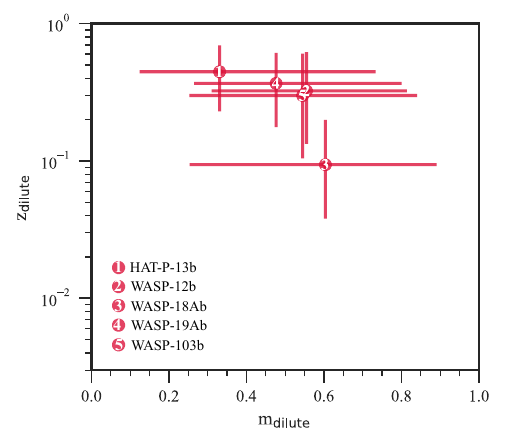}
    \caption{The interior structure parameters of the planets in the sample. In the left panel we show the core mass fraction and atmospheric metallicity. The background lines represent equal bulk metallicity ($Z_\mathrm{planet}$) for the homogeneous model. The right panel shows the dilute core parameters $m_\mathrm{dilute}$ and $Z_\mathrm{dilute}$. Jupiter's and Saturn's most likely core mass, here assumed to be the metal content in the compact + dilute core \citep{Sur-2024-2}, and atmospheric metallicity, calculated from the elemental O/H abundance for Jupiter \citep{Bjoraker_2018} the elemental C/H abundance for Saturn \citep{Fletcher_2009}, are shown in green and yellow  for comparison. }
    \label{fig:interiorparams_planets}
\end{figure*}

\begin{figure}
    \centering
    \includegraphics[width=\columnwidth]{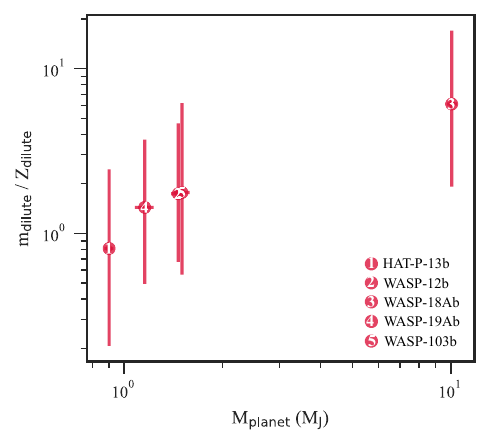}
    \caption{The ratio between the dilute core parameters $m_\mathrm{dilute}$ and $Z_\mathrm{dilute}$ and mass of the planets in the sample. A higher ratio indicates a more extended metal-poor dilute core, while a lower ratio indicates a more compact metal-rich dilute core.}
    \label{fig:mdilutezdilute_mass}
\end{figure}

Figure \ref{fig:interiorparams_planets} presents the retrieved interior structure parameters for the five planets. The exact values can be found in Table \ref{tab:results_HATP13b}-\ref{tab:results_WASP103b} in Appendix \ref{ap:tables_results}. For the homogeneous model, the interior structure and metal content is solely determined by the core mass fraction and the atmospheric metallicity, shown in the left panel. Therefore, we can draw constant metallicity lines in this parameter space. All planets exhibit similar bulk metallicities of approximately 30\% in mass, although the distribution of metals within the interior differs. HAT-P-13b and WASP-19Ab have a lower core mass fraction and a higher atmospheric metallicity, while WASP-12b, WASP-18Ab and WASP-103b have a higher core mass fraction with less metals in the envelope. However, this trend should be interpreted cautiously, as it likely reflects the uncertainties in the atmospheric metallicity and Love number measurements. In Section \ref{sec:test_planet}, we established that to obtain accurate constraints on the interior properties of hot Jupiters we need a constraint on the atmospheric metallicity and a Love number precision better than approximately $40$\%. As the Love number measurements of WASP-12b, WASP-18Ab and WASP-103b exceed this threshold, it is likely that their large uncertainties in the Love number are leading to an overestimation of the retrieved core mass. Similarly, because HAT-P-13b lacks an atmospheric metallicity measurement, its atmospheric metallicity, and therefore also its bulk metallicity, is less constrained and likely overestimated.

For the inhomogeneous model, the dilute core also contributes to the bulk metallicity and metal distribution. The right panel in Figure \ref{fig:interiorparams_planets} shows the corresponding dilute core parameters for the five planets. The planet dilute core properties generally follow a similar trend as observed in the left panel for the homogeneous model: a lower metal content in the dilute core is associated with a more extended dilute core, while a higher metal content corresponds to a smaller dilute core. This leads, analogously to the homogeneous case, to similar bulk metallicities across the five planets. The similarity between the left and right panel may be influenced by the pre-set physical priors ($m_\mathrm{dilute}>m_\mathrm{core}$, $Z_\mathrm{dilute}>Z_\mathrm{atm}$) and are therefore also likely affected by uncertainties in the atmospheric metallicity and Love number measurement. Interestingly, this trend correlates with the masses of the planets, as is more clearly shown in Figure \ref{fig:mdilutezdilute_mass}. We find a lower $m_\mathrm{dilute}$/$Z_\mathrm{dilute}$ ratio, which indicates more compact metal-rich dilute cores, for lower-mass planets, while the higher-mass planets have higher $m_\mathrm{dilute}$/$Z_\mathrm{dilute}$ ratios, indicating more extended metal-poor dilute cores.

\subsubsection{HAT-P-13b}
HAT-P-13b's interior has been modelled extensively in numerous previous works \citep[e.g.][]{Batygin2009_firstkeccentricity, Kramm2012ConstrainingHAT-P-13b, Buhler_2016, Hardy_2017}. We compare our results with \citet{Buhler_2016}, who derived the Love number measurement that we are using in this work. There are a few differences in their approach. They use evolution models with an older equation of state for hydrogen and helium \citep{EoS_helium}, assume a solar composition atmosphere and only use dissipation rates between 0.05 to 0.5\% of the incoming flux. They find a most probable core mass of 11 $M_\oplus$, a 2-sigma upper limit of 25 $M_\oplus$ and a 3-sigma upper limit of 47 $M_\oplus$. We find a 2-sigma and 3-sigma upper limit of 17.5 $M_\oplus$ and 38 $M_\oplus$, which are slightly lower than what \citet{Buhler_2016} find. In contrast to their fixed solar metallicity atmosphere, (and given the lack of measurements) we keep this parameter free in the retrieval. This results in more metals in the atmosphere and could potentially reduce the amount of metals in the core of the planet in our models. Future atmospheric determinations of this planet with JWST will shed more light on the core mass determination of this planet. 

\subsubsection{WASP-19Ab}
Among the planets in our sample, WASP-19Ab stands out with the most precise Love number determination and a well-constrained atmospheric metallicity, providing the strongest constraints on its interior. To the best of our knowledge, its interior has not been modelled before. 
Thanks to this high-quality data, we confidently confirm the presence of a core in WASP-19Ab. While we cannot yet distinguish whether the planet has a homogeneous envelope with a compact core or requires a dilute core, the existence of a core is robust. 
Using a homogeneous model, we derive a core mass of $79^{+21}_{-18} M_\oplus$, corresponding to a core mass fraction of approximately $0.21^{+0.05}_{-0.04}$. This is particularly significant, as we can only place upper limits on core masses for other planets in our sample.
When a dilute core is considered, we find a compact core mass of $55^{+25}_{-29}$ M$_\oplus$, but the distribution extends to zero. However, the probability of the planet lacking a core entirely is negligibly small. Specifically, the likelihood that both the core mass fraction and the extent of the dilute core region are below 0.1 is only 0.02\%. This effectively rules out a coreless interior for this planet and puts firm constraints on what to expect for the distribution of metals inside this giant.

\subsection{Bulk metal content}

\begin{figure}
    \centering
    \includegraphics[width=\columnwidth]{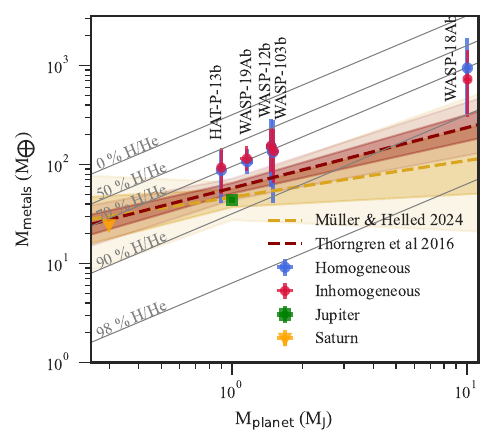}
    \caption{The bulk metal mass as a function of planet mass for the five planets in the sample. The background lines represent lines of equal bulk metallicity ($Z_\mathrm{planet}$). We compare our results to the mass-metallicity trends derived by \citet{Thorngren_2016} and \citet{Muller_2024_2}, with the shaded regions indicating the 1- and 2-sigma spread. The most likely metal masses for Jupiter and Saturn are shown in green and yellow for comparison \citep{Sur-2024-2}.}
    \label{fig:bulk_metals_mplanet_planets}
\end{figure}

\begin{figure}
    \centering
    \includegraphics[width=\columnwidth]{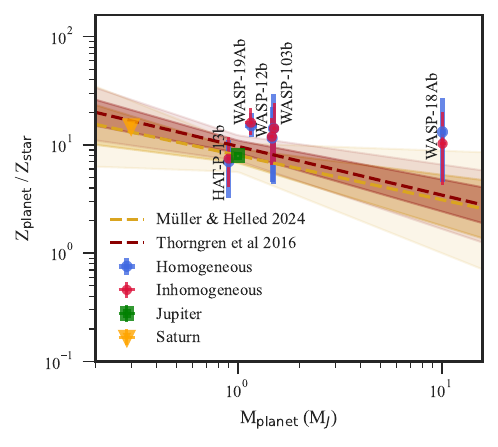}
    \caption{The bulk metal enrichment for the five planets as a function of planet mass. We compare these results to the mass-metallicity trends derived by \citet{Thorngren_2016} and \citet{Muller_2024_2}, with the shaded regions indicating the 1- and 2-sigma spread. The most likely metal enrichments for Jupiter and Saturn are shown in green and yellow for comparison \citep{Sur-2024-2}.}
    \label{fig:metal_enrichment_mplanet_planets}
\end{figure}

\begin{figure}
    \centering
    \includegraphics[width=\columnwidth]{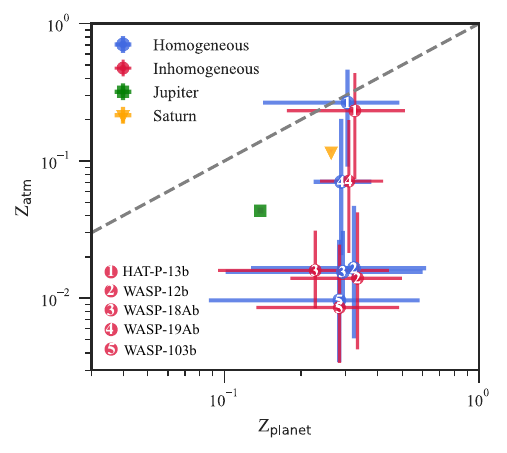}
    \caption{The bulk metallicity and atmospheric metallicity of the five planets in the sample. Jupiter and Saturn are shown for comparison with bulk metallicities from \citet{Sur-2024-2} and atmospheric metallicities calculated from the O/H (for Jupiter) and C/H (for Saturn) atmospheric abundances \citep{Bjoraker_2018, Fletcher_2009}. The dashed line represents equal metal mass fraction in the atmosphere and the rest of the planet, indicating a fully mixed planet.}
    \label{fig:bulkmetal_vs_atmmetal}
\end{figure}

Figure \ref{fig:bulk_metals_mplanet_planets} shows the bulk metal mass of the planets as a function of their mass. A relation between the metal content and planet mass has been proposed in various previous studies \citep[e.g.][]{Miller_2011, Thorngren_2016, Muller_2024_2}. We compare our results with the most recent trends, that were obtained by modelling warm gas giant exoplanets with evolutionary interior models, as opposed to our results obtained with static model calculations and considering also inflated hot-Jupiters. We generally find higher metal masses for the five planets in our sample. However, our results do agree within one sigma for HAT-P-13b, WASP-12b and WASP-103b, and within two sigma for WASP-18Ab and WASP-19Ab. When extending the analysis to bulk metal enrichment as a function of mass (see Figure \ref{fig:metal_enrichment_mplanet_planets}), for which previous works found a more pronounced relationship, we find that HAT-P-13b, WASP-12b, WASP-103b and WASP-18Ab agree within one sigma, while WASP-19Ab again agrees within two sigma.

Overall, the inhomogeneous model retrieves very similar bulk metallicities compared to the homogeneous model. For most planets the values are slightly higher when using the inhomogeneous model compared to the homogeneous model, but this difference is small compared to the dispersion in the retrieved values (see Figure \ref{fig:bulk_metals_mplanet_planets} and \ref{fig:metal_enrichment_mplanet_planets}). For WASP-18Ab, the difference is larger than for the other planets. In this case, the homogeneous model retrieves a higher metallicity than the inhomogeneous model, although the one-sigma uncertainties still overlap. This higher metallicity for the homogeneous model, in contrast with the lower metallicities retrieved for the other planets, can be explained by the densities in the dilute core region. For WASP-18Ab, the densities in the dilute core are, relative to the other planets, higher compared to the rest of the envelope. As a result, the inhomogeneous model is denser than a homogeneous model with the same amount of metals and needs fewer metals to explain the radius and mass of the planet. 

It is relevant to compare bulk metallicities with atmospheric metallicities, for example to gain information about interior-atmosphere interactions and establish how well-mixed the planet's interior is. We compare the retrieved atmospheric metallicities with the bulk metallicities in Figure \ref{fig:bulkmetal_vs_atmmetal}. Overall, the planets do not follow a trend and rather show a large dispersion of $Z_{\mathrm{atm}}$, indicating that the atmospheric metallicity is not necessarily indicative of the total planetary bulk metallicity. For the higher mass planets with low atmospheric metallicities, such as WASP-12b, WASP-103 and WASP-18Ab, a higher amount of metals in the (dilute) core is necessary to explain their masses and radii.

\subsection{Heating efficiency}

\begin{figure}
    \centering
    \includegraphics[width=\columnwidth]{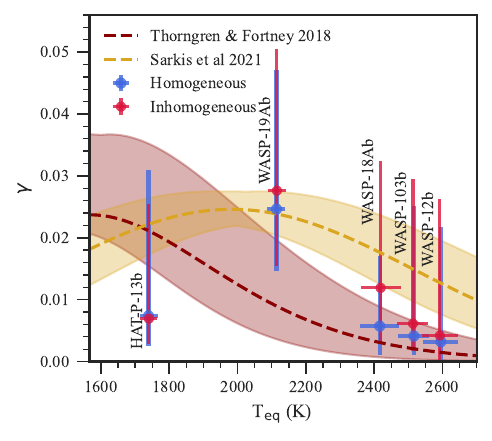}
    \caption{The heating efficiency fraction and the equilibrium temperatures of the planets with Love number measurements. We compare our results to the Heating Efficiency Equilibrium Temperature (HEET) distributions derived by \citet{thorngren2018} and \citet{Sarkis2021} with the shaded region indicating the 1-sigma spread of the distributions.}
    \label{fig:heating_planets}
\end{figure}

Figure \ref{fig:heating_planets} shows the derived heating efficiencies as a function of equilibrium temperature for the five planets. We compare our findings to general Heating Efficiency Equilibrium Temperature (HEET) distributions derived from a sample of hot Jupiters by \citet{thorngren2018} and \citet{Sarkis2021}. With slightly different approaches (\citet{thorngren2018} use evolution models and \citet{Sarkis2021} use static models), they derive for the population an underlaying Gaussian distribution, with peaks around 1566 K and 1860 K, respectively. Although we derive heating efficiencies for individual planets, resulting in less constrained values, we find that the values for all planets in our sample are within the 1-sigma range of their results.
\section{Discussion} \label{sec:discussion}

\subsection{Non-linear tidal response} \label{sec:nonlineartidalresponse}

There are certain limitations in our approach to calculate the Love number from a given density profile (see Section \ref{sec:methods_lovenumber}). One key limitation is that we assume a linear tidal response and neglect non-linear contributions that may arise, for example, due to rotation. \citet{Wahl_2021} demonstrated, using the more accurate concentric MacLaurin spheroid method, that this approximation only holds for hot Jupiters with small rotational parameters ($q_0<<0.01$)\footnote{The rotational parameter $q_0$ is defined as the ratio between the centrifugal and gravitational forces.}. When the rotational parameter exceeds this threshold, non-linear effects become significant. In our sample, two planets affected by these non-linear effects - WASP-12b and WASP-103b - overlap with those studied by \citet{Wahl_2021}. However, the observational uncertainties of the Love numbers of these planets are larger than the potential underestimation of the Love number caused by ignoring non-linear tidal effects. Therefore, we conclude that this does not significantly impact our results for WASP-12b and WASP-103b. 

There is one planet in our sample, WASP-19Ab, that was not included in \citet{Wahl_2021}'s analysis. Assuming synchronous rotation, the rotational parameter for WASP-19Ab is approximately $0.06$, which is too large for non-linear effects to be ignored. This rotational parameter is similar to that of WASP-12b, for which \citet{Wahl_2021} estimated that neglecting non-linear effects would result in an underestimation of the Love number by around $19$\%. Given that the observational uncertainty of WASP-19Ab's Love number is only about $12$\%, the potential underestimation in the Love number could impact our results. Therefore, we recommend that future studies account for non-linear tidal effects when modelling the interior of WASP-19Ab for greater accuracy.

\subsection{Future outlook}

In both the experiments with the test planet (see Section \ref{sec:test_planet}) and the retrievals from our exoplanet sample (see Section \ref{sec:real_planets}), we found that obtaining accurate constraints on the core mass fraction requires a precise measurement of the Love number, with an uncertainty of less than 40\%. Currently, only two out of the five exoplanets with measured Love numbers meet this precision threshold: HAT-P-13b, with a precision of 21\% \citep{Buhler_2016}, and WASP-19Ab, with a precision of 12\% \citep{Bernabo_2024}. The exoplanets with the largest uncertainties, WASP-12b and WASP-103b, have Love numbers measured by directly modelling their tidal deformation via light curve observations \citep{Akinsanmi-2024-2, Barros_2022}.
 
\citet{Akinsanmi_2024} investigated the detectability of the tidal deformation in the phase and transit light curves for current and upcoming space missions (JWST, PLATO, HST, TESS, CHEOPS), using WASP-12b as a case study. They expect the best results from JWST. With a single NIRSpec-PRISM JWST phase curve, they expect a Love number detection of $h_{2}=1.52^{+0.094}_{-0.087}$\footnote{Under hydrostatic equilibrium $h_{2} = 1 + k_{2}$.}, corresponding to Love number precision of approximately $17$\%. A single transit curve would result in a Love number detection of $h_{2}=1.55^{+0.32}_{-0.37}$, resulting in a precision of 63\%. Other instruments perform similarly or worse than the JWST transit curve fit. In summary, out of the configurations that \citet{Akinsanmi_2024} test, only a single phase curve observations with JWST would result in a Love number precision below our required threshold. For other predictions towards future observations,  we refer the interested reader to \citet{Hellard_2019}.

The future prospects for this field are highly promising, as upcoming JWST programmes will provide more precise Love numbers and better atmospheric data for the planets discussed in this study and other planets. Together, these new observational constraints will drive progress toward a more detailed characterization of the interiors of giant exoplanets.
\section{Conclusions}

In this study, we examine how new observational constraints, such as the Love number and atmospheric metallicity, affect our ability to infer the interior properties of hot Jupiters. We use both homogeneous and inhomogeneous models and account for observational uncertainties, including those associated with mass, radius, equilibrium temperature and inflation by applying a Bayesian framework. 

Our results demonstrate that combining precise Love number measurements with atmospheric metallicity observations significantly improves constraints on planetary interiors. Using both constraints, with a Love number precision better than 40\% for the homogeneous model and 15\% for the inhomogeneous model, we can constrain the bulk metallicity within one sigma of the true value. The Love number alone provides a strong constraint on the core mass for both the homogeneous and inhomogeneous model, consistent with previous work \citep[e.g.][]{Kramm2011} but newly established here for the inhomogeneous case. Additionally, the Love number improves constraints on heating efficiency, which could help understanding the mechanisms driving hot Jupiter inflation. In the inhomogeneous model, the Love number also helps constrain the extent of the dilute core, although with broad dispersion, and more effectively constrains the metal content in the dilute core. 

We apply our retrieval framework to five planets with Love number measurements: HAT-P-13b, WASP-12b, WASP-18Ab, WASP-19Ab and WASP-103b. Current Love number measurements vary significantly in precision, with some uncertainties being too large to provide meaningful constraints (e.g. WASP-12b with 85\% \citep{Akinsanmi_2024} and WASP-103b with 83\% \citep{Barros_2022}). Our analysis shows that Love number measurements with uncertainties exceeding 40\% do not substantially improve interior constraints. Of the planets studied, only HAT-P-13b and WASP-19Ab meet the required threshold, leading to significantly better core mass constraints. For WASP-19Ab we can even confirm the existence of a core, though both a compact or diluted core scenario agree with the observations. Of the modelled planets, WASP-19Ab also provides the best bulk metallicity constraint due to the availability of both a precise Love number and atmospheric metallicity measurement, whereas HAT-P-13b lacks the latter. 

These findings highlight the critical role of high-precision Love number and atmospheric metallicity measurements to further improve hot Jupiter interior constraints and shows the importance of interior modelling to plan for future observations. Upcoming JWST observations offer improvements for both precise Love numbers and atmospheric metallicity measurements, through full phase curve observations \citep{Akinsanmi-2024-2} and high-precision spectroscopy. The synergy between these measurements will provide a more complete picture of hot Jupiter interiors, bringing us closer to resolving key questions about their composition and structure.


\section*{Acknowledgements}

This work has been done thanks to support from the European Research Council (ERC) under the European Union’s Horizon 2020 research and innovation programme (grant agreement no. 101088557, N-GINE).
This research made use of NASA’s Astrophysics Data System, \textsc{NumPy} \citep{numpy}, \textsc{MATPLOTLIB}, a \textsc{PYTHON} library for publication quality graphics \citep{matplotlib}, \textsc{SciPy} \citep{2020SciPy-NMeth}.

\section*{Data Availability}

The data underlying this article will be shared on reasonable request to the corresponding author.




\bibliographystyle{mnras}
\bibliography{references} 




\appendix
\section{Metallicity conversion} \label{ap:metallicity}

In our interior models the composition is defined by mass fractions of hydrogen (X), helium (Y) and metals (Z). By definition these fractions sum up to 1.

\begin{equation}
    X + Y + Z \equiv 1
\end{equation}

The heavy element fraction $Z$ can thus be written as:
\begin{equation}
    Z = \frac{M_Z}{M_X+M_Y+M_Z}. \label{eq:mass_fraction_Z}
\end{equation}

We write $M_Z$ as the sum of masses of different metal species $i$ in the interior and convert every mass $M$ to the number density $N$ using the molecular weight $\mu$.

\begin{equation}
    Z = \frac{\sum_i M_i}{M_{\mathrm{H}} + M_{\mathrm{He}} + \sum_i M_i} = \frac{\sum_i \mu_i N_i}{\mu_{\mathrm{H}} N_{\mathrm{H}} + \mu_{\mathrm{He}} N_{\mathrm{He}} + \sum_i \mu_i N_i}
\end{equation}
 
The observed metallicity is often given relative to solar in dex: $[M/H]$. This can be converted to relative number densities by assuming a set of solar abundances.

\begin{equation}
    10^{[\mathrm{M/H}]} = \frac{10^{(\mathrm{M/H})_{\mathrm{planet}}}}{10^{(\mathrm{M/H})_\odot}} = \frac{(N_\mathrm{M}/N_\mathrm{H})_{\mathrm{planet}}}{(N_\mathrm{M}/N_\mathrm{H})_\odot}
\end{equation}

Assuming that every metal is similarly enriched or depleted compared to solar, we can substitute this in Equation \ref{eq:mass_fraction_Z}. Additionally, we assume that the hydrogen-to-helium ratio is proto-solar and can finally calculate the mass fraction $Z$ using only a set of solar abundances and the metal enrichment/depletion relative to solar.

\begin{equation}
    Z =  \frac{10^{[\mathrm{M/H}]} \sum_i \mu_i (N_i/N_\mathrm{H})_\odot}{\mu_\mathrm{H} + \mu_\mathrm{He} (N_{\mathrm{He}}/N_{\mathrm{H}})_\odot + 10^{[\mathrm{M/H}]} \sum_i \mu_i (N_i/N_\mathrm{H})_\odot} 
\end{equation}
\section{Internal luminosity prior}\label{ap:internal_luminosity}

Sections \ref{sec:test_planet} and \ref{sec:real_planets} only show results, where the prior on parameters associated with the internal luminosity, i.e. the luminosity a non-inflated planet would have, $L_\mathrm{grav}$, and the heating efficiency fraction, $\gamma$, are set to log-uniform. Here, we compare those results with the results obtained with the linear-uniform prior. 

\subsection{Test planet}

\begin{figure*}
    \centering
    \includegraphics[width=\textwidth]{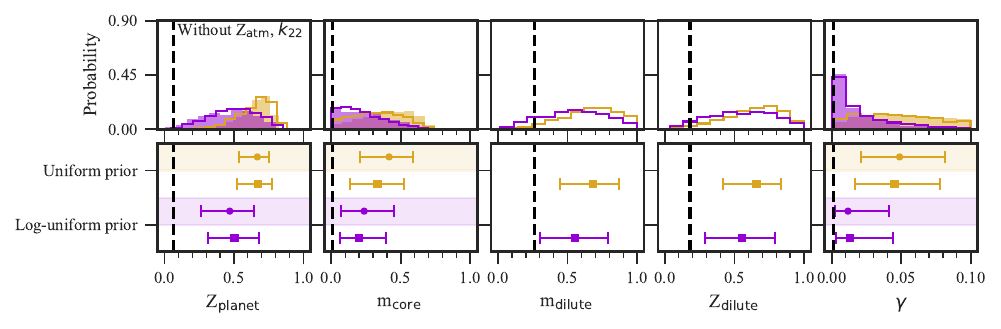}
    \includegraphics[width=\textwidth]{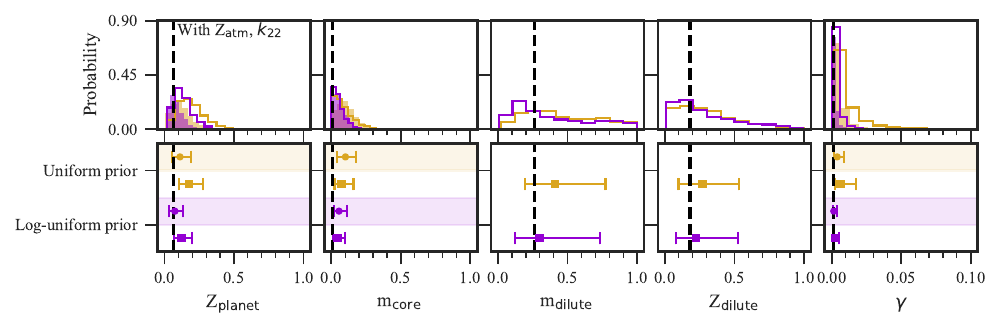}
    \caption{The resulting posterior distributions (upper panels) and confidence intervals (lower panels) for retrievals with a linear-uniform prior (yellow) and log-uniform prior (purple) for the internal luminosity parameters. The upper figure shows the results of the test planet without using measurements on the atmospheric metallicity and Love number and the lower figure shows results for the test planet using those extra measurements (see Sec. \ref{sec:test_planet}). The shaded regions show the homogeneous model and unshaded regions the inhomogeneous model.}
    \label{fig:test_planet_Lintcomparison}
\end{figure*}

 In Figure \ref{fig:test_planet_Lintcomparison} we show the difference in posteriors for two of the test planet scenarios from Section \ref{sec:testplanet_adding_z_k}, one where an atmospheric metallicity and Love number measurement are included and one without them. With both a constrained Love number and atmospheric metallicity (shown in the lower panel), we find similar results for both priors. However, without an atmospheric metallicity and Love number measurement (shown in the upper panel), the posteriors are different. The log-uniform prior favours smaller internal luminosities and heating efficiencies. Therefore, because for this test planet the true heating efficiency is relatively small (0.001), the log-uniform prior finds a better solution than the linear-uniform prior. 

\subsection{Sample of exoplanets}

\begin{figure*}
    \centering
    \includegraphics[width=\textwidth]{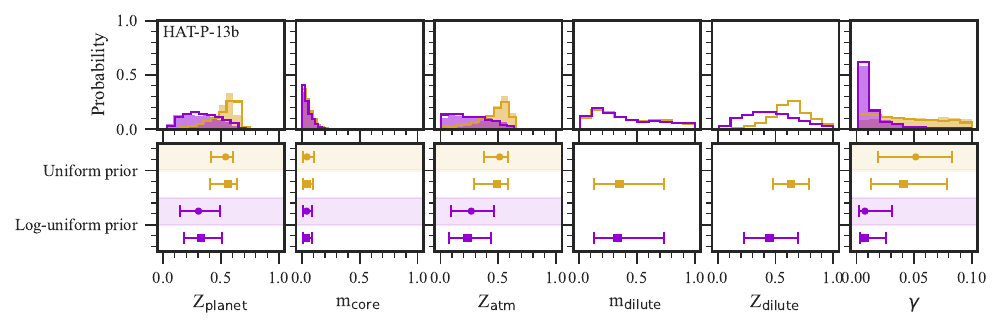}
    
    \caption{The resulting posterior distributions (upper panels) and confidence intervals (lower panels) for retrievals for HAT-P-13b with a linear-uniform prior (yellow) and log-uniform prior (purple) for the internal luminosity parameters. The shaded regions show the homogeneous model and the unshaded regions the inhomogeneous model.}
    \label{fig:real_planets_Lintcomparison_HAT-P-13b}
\end{figure*}

\begin{figure*}
    \centering
    \includegraphics[width=\textwidth]{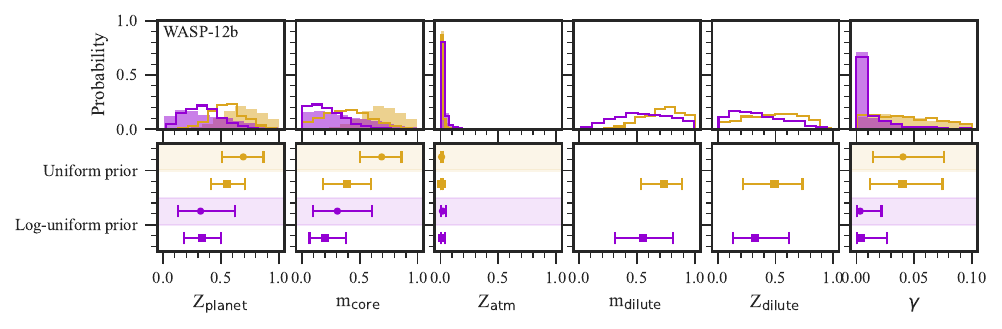}
    \caption{The resulting posterior distributions (upper panels) and confidence intervals (lower panels) for retrievals for WASP-12b with a linear-uniform prior (yellow) and log-uniform prior (purple) for the internal luminosity parameters. The shaded regions show the homogeneous model and the unshaded regions the inhomogeneous model.}
    \label{fig:real_planets_Lintcomparison_WASP-12b}
\end{figure*}

\begin{figure*}
    \centering
    \includegraphics[width=\textwidth]{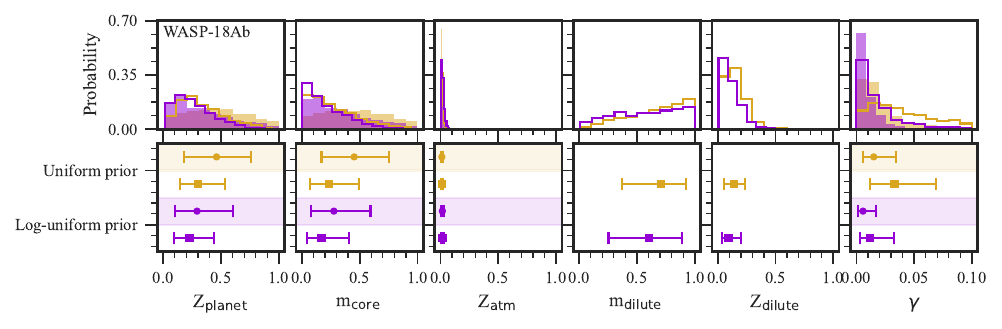}
    \caption{The resulting posterior distributions (upper panels) and confidence intervals (lower panels) for retrievals for WASP-18Ab with a linear-uniform prior (yellow) and log-uniform prior (purple) for the internal luminosity parameters. The shaded regions show the homogeneous model and the unshaded regions the inhomogeneous model.}
    \label{fig:real_planets_Lintcomparison_WASP-18Ab}
\end{figure*}

\begin{figure*}
    \centering
    \includegraphics[width=\textwidth]{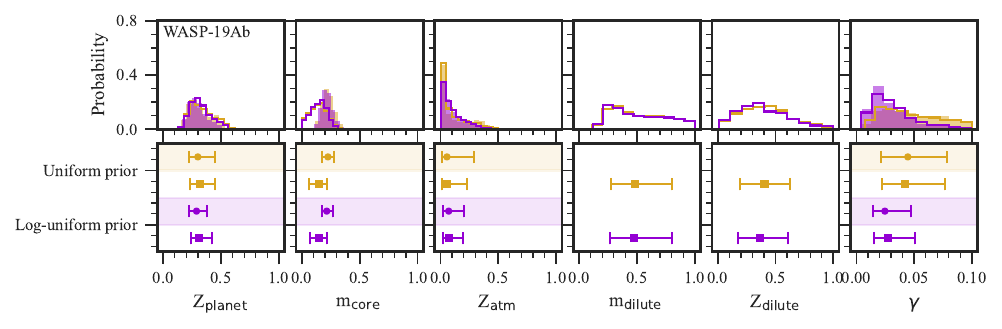}
    \caption{The resulting posterior distributions (upper panels) and confidence intervals (lower panels) for retrievals for WASP-19Ab with a linear-uniform prior (yellow) and log-uniform prior (purple) for the internal luminosity parameters. The shaded regions show the homogeneous model and the unshaded regions the inhomogeneous model.}
    \label{fig:real_planets_Lintcomparison_WASP-19Ab}
\end{figure*}

\begin{figure*}
    \centering
    \includegraphics[width=\textwidth]{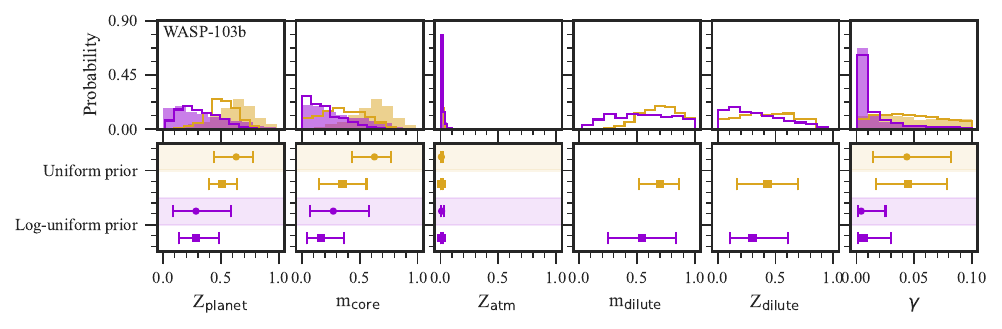}
    \caption{The resulting posterior distributions (upper panels) and confidence intervals (lower panels) for retrievals for WASP-103b with a linear-uniform prior (yellow) and log-uniform prior (purple) for the internal luminosity parameters. The shaded regions show the homogeneous model and the unshaded regions the inhomogeneous model.}
    \label{fig:real_planets_Lintcomparison_WASP-103b}
\end{figure*}

In Figures \ref{fig:real_planets_Lintcomparison_HAT-P-13b}, \ref{fig:real_planets_Lintcomparison_WASP-12b}, \ref{fig:real_planets_Lintcomparison_WASP-18Ab}, \ref{fig:real_planets_Lintcomparison_WASP-19Ab} and \ref{fig:real_planets_Lintcomparison_WASP-103b} we the results obtained with both priors for the planets in the sample from Section \ref{sec:real_planets}. The results show good agreement between the two priors for WASP-19Ab and WASP-18Ab, two planets with relatively precise Love number measurements and available atmospheric metallicity data. However, using the linear-uniform prior, we observe larger inferred core masses for WASP-12b and WASP-103b, as well as higher bulk metallicities for WASP-12b, WASP-103b, and HAT-P-13b, due to the higher internal temperatures in the retrieval. The first two planets have highly uncertain Love number measurements, while HAT-P-13b lacks atmospheric metallicity data.

\section{Planet results} \label{ap:tables_results}

\begin{table*}
    \centering
    \begin{tabular}{l|rrrr}
    \hline
 & Homogeneous LU & Inhomogeneous LU & Homogeneous U & Inhomogeneous U \\ \hline

Z$_\mathrm{planet}$& $0.30^{+0.18}_{-0.16}$& $0.33^{+0.19}_{-0.15}$& $0.55^{+0.06}_{-0.13}$& $0.56^{+0.08}_{-0.18}$\\
M$_\mathrm{metals}$ (M$_{\oplus}$)& $87.42^{+51.84}_{-46.35}$& $93.11^{+53.24}_{-42.28}$& $156.30^{+16.92}_{-37.65}$& $159.99^{+23.41}_{-50.68}$\\
m$_{\mathrm{core}}$& <0.06& <0.06& <0.06& <0.06\\
M$_\mathrm{core}$ (M$_{\oplus}$)& <17.48& <16.83& <18.46& <18.64\\
m$_\mathrm{dilute}$& - & $0.33^{+0.40}_{-0.21}$& - & $0.33^{+0.42}_{-0.20}$\\
Z$_\mathrm{dilute}$& - & $0.45^{+0.25}_{-0.22}$& - & $0.63^{+0.17}_{-0.17}$\\
Z$_\mathrm{atm}$& $0.27^{+0.20}_{-0.18}$& $0.23^{+0.20}_{-0.16}$& $0.52^{+0.07}_{-0.15}$& $0.49^{+0.09}_{-0.21}$\\
$\gamma$& $0.01^{+0.02}_{-0.00}$& $0.01^{+0.02}_{-0.00}$& $0.05^{+0.03}_{-0.04}$& $0.04^{+0.04}_{-0.03}$\\
\hline
    \end{tabular}
    \caption{The retrieved interior properties of HAT-P-13b. From left to right the columns show the results obtained with the homogeneous model and a log-uniform prior on the internal luminosity parameters, the inhomogeneous model and a log-uniform prior on the internal luminosity parameters, the homogeneous model and a linear-uniform prior on the internal luminosity parameters, and the inhomogeneous model and a linear-uniform prior on the internal luminosity parameters. }
    \label{tab:results_HATP13b}
\end{table*}

\begin{table*}
    \centering
    \begin{tabular}{l|rrrr}
    \hline
 & Homogeneous LU & Inhomogeneous LU & Homogeneous U & Inhomogeneous U \\ \hline
Z$_\mathrm{planet}$& $0.32^{+0.30}_{-0.20}$& $0.33^{+0.16}_{-0.15}$& $0.69^{+0.18}_{-0.18}$& $0.55^{+0.15}_{-0.14}$\\
M$_\mathrm{metals}$ (M$_{\oplus}$)& $150.60^{+138.81}_{-90.89}$& $154.46^{+79.75}_{-70.95}$& $323.25^{+80.72}_{-85.32}$& $256.70^{+71.68}_{-63.94}$\\
m$_{\mathrm{core}}$& <0.44& <0.28& <0.77& <0.49\\
M$_\mathrm{core}$ (M$_{\oplus}$)& <207.22& <129.05& <360.84& <226.32\\
m$_\mathrm{dilute}$& - & $0.56^{+0.26}_{-0.24}$& - & $0.73^{+0.16}_{-0.20}$\\
Z$_\mathrm{dilute}$& - & $0.32^{+0.30}_{-0.19}$& - & $0.49^{+0.24}_{-0.27}$\\
Z$_\mathrm{atm}$& $0.02^{+0.03}_{-0.01}$& $0.01^{+0.03}_{-0.01}$& $0.01^{+0.01}_{-0.01}$& $0.01^{+0.01}_{-0.00}$\\
$\gamma$& $0.00^{+0.02}_{-0.00}$& $0.00^{+0.02}_{-0.00}$& $0.04^{+0.04}_{-0.03}$& $0.04^{+0.03}_{-0.03}$\\
\hline
    \end{tabular}
    \caption{The retrieved interior properties of WASP-12b. From left to right the columns show the results obtained with the homogeneous model and a log-uniform prior on the internal luminosity parameters, the inhomogeneous model and a log-uniform prior on the internal luminosity parameters, the homogeneous model and a linear-uniform prior on the internal luminosity parameters, and the inhomogeneous model and a linear-uniform prior on the internal luminosity parameters.}
    \label{tab:results_WASP12b}
\end{table*}

\begin{table*}
    \centering
    \begin{tabular}{l|rrrr}
    \hline
 & Homogeneous LU & Inhomogeneous LU & Homogeneous U & Inhomogeneous U \\ \hline
Z$_\mathrm{planet}$& $0.29^{+0.31}_{-0.19}$& $0.23^{+0.22}_{-0.13}$& $0.46^{+0.30}_{-0.28}$& $0.30^{+0.24}_{-0.15}$\\
M$_\mathrm{metals}$ (M$_{\oplus}$)& $942.89^{+960.36}_{-621.69}$& $727.17^{+695.05}_{-424.08}$& $1498.25^{+952.09}_{-904.25}$& $969.86^{+764.16}_{-485.83}$\\
m$_{\mathrm{core}}$& <0.41& <0.27& <0.60& <0.35\\
M$_\mathrm{core}$ (M$_{\oplus}$)& <1321.84& <873.20& <1937.78& <1127.46\\
m$_\mathrm{dilute}$& - & $0.60^{+0.29}_{-0.35}$& - & $0.71^{+0.21}_{-0.34}$\\
Z$_\mathrm{dilute}$& - & $0.09^{+0.10}_{-0.06}$& - & $0.14^{+0.09}_{-0.09}$\\
Z$_\mathrm{atm}$& $0.02^{+0.02}_{-0.01}$& $0.02^{+0.02}_{-0.01}$& $0.01^{+0.01}_{-0.00}$& $0.01^{+0.01}_{-0.01}$\\
$\gamma$& $0.01^{+0.01}_{-0.00}$& $0.01^{+0.02}_{-0.01}$& $0.01^{+0.02}_{-0.01}$& $0.03^{+0.04}_{-0.02}$\\
\hline
    \end{tabular}
    \caption{The retrieved interior properties of WASP-18Ab. From left to right the columns show the results obtained with the homogeneous model and a log-uniform prior on the internal luminosity parameters, the inhomogeneous model and a log-uniform prior on the internal luminosity parameters, the homogeneous model and a linear-uniform prior on the internal luminosity parameters, and the inhomogeneous model and a linear-uniform prior on the internal luminosity parameters.}
    \label{tab:results_WASP18Ab}
\end{table*}

\begin{table*}
    \centering
    \begin{tabular}{l|rrrr}
    \hline
 & Homogeneous LU & Inhomogeneous LU & Homogeneous U & Inhomogeneous U \\ \hline
Z$_\mathrm{planet}$& $0.29^{+0.09}_{-0.06}$& $0.31^{+0.11}_{-0.07}$& $0.30^{+0.15}_{-0.07}$& $0.32^{+0.13}_{-0.08}$\\
M$_\mathrm{metals}$ (M$_{\oplus}$)& $107.04^{+36.49}_{-26.44}$& $114.74^{+40.53}_{-28.90}$& $112.20^{+53.95}_{-30.33}$& $118.38^{+51.75}_{-32.86}$\\
m$_{\mathrm{core}}$& $0.21^{+0.05}_{-0.04}$& $0.15^{+0.06}_{-0.08}$& $0.23^{+0.05}_{-0.05}$& $0.15^{+0.07}_{-0.08}$\\
M$_\mathrm{core}$ (M$_{\oplus}$)& $79.25^{+21.30}_{-18.22}$& $54.53^{+25.22}_{-28.60}$& $82.92^{+22.64}_{-19.98}$& $53.20^{+28.80}_{-30.21}$\\
m$_\mathrm{dilute}$& - & $0.48^{+0.32}_{-0.21}$& - & $0.49^{+0.32}_{-0.21}$\\
Z$_\mathrm{dilute}$& - & $0.37^{+0.24}_{-0.19}$& - & $0.41^{+0.22}_{-0.22}$\\
Z$_\mathrm{atm}$& $0.07^{+0.13}_{-0.05}$& $0.07^{+0.13}_{-0.05}$& $0.06^{+0.23}_{-0.05}$& $0.05^{+0.18}_{-0.04}$\\
$\gamma$& $0.02^{+0.02}_{-0.01}$& $0.03^{+0.02}_{-0.01}$& $0.04^{+0.03}_{-0.02}$& $0.04^{+0.03}_{-0.02}$\\
\hline
    \end{tabular}
    \caption{The retrieved interior properties of WASP-19Ab. From left to right the columns show the results obtained with the homogeneous model and a log-uniform prior on the internal luminosity parameters, the inhomogeneous model and a log-uniform prior on the internal luminosity parameters, the homogeneous model and a linear-uniform prior on the internal luminosity parameters, and the inhomogeneous model and a linear-uniform prior on the internal luminosity parameters.}
    \label{tab:results_WASP19Ab}
\end{table*}

\begin{table*}
    \centering
    \begin{tabular}{l|rrrr}
    \hline
 & Homogeneous LU & Inhomogeneous LU & Homogeneous U & Inhomogeneous U \\ \hline
Z$_\mathrm{planet}$& $0.28^{+0.30}_{-0.20}$& $0.28^{+0.20}_{-0.15}$& $0.63^{+0.15}_{-0.19}$& $0.51^{+0.13}_{-0.12}$\\
M$_\mathrm{metals}$ (M$_{\oplus}$)& $133.01^{+146.21}_{-92.09}$& $136.58^{+94.09}_{-71.97}$& $297.56^{+75.47}_{-91.71}$& $240.11^{+63.37}_{-55.51}$\\
m$_{\mathrm{core}}$& <0.41& <0.24& <0.70& <0.46\\
M$_\mathrm{core}$ (M$_{\oplus}$)& <196.52& <115.01& <332.41& <214.29\\
m$_\mathrm{dilute}$& - & $0.55^{+0.30}_{-0.29}$& - & $0.70^{+0.16}_{-0.18}$\\
Z$_\mathrm{dilute}$& - & $0.30^{+0.30}_{-0.20}$& - & $0.43^{+0.26}_{-0.27}$\\
Z$_\mathrm{atm}$& $0.01^{+0.02}_{-0.01}$& $0.01^{+0.02}_{-0.01}$& $0.01^{+0.01}_{-0.00}$& $0.01^{+0.01}_{-0.00}$\\
$\gamma$& $0.00^{+0.02}_{-0.00}$& $0.01^{+0.02}_{-0.00}$& $0.04^{+0.04}_{-0.03}$& $0.04^{+0.03}_{-0.03}$\\
\hline
    \end{tabular}
    \caption{The retrieved interior properties of WASP-103b. From left to right the columns show the results obtained with the homogeneous model and a log-uniform prior on the internal luminosity parameters, the inhomogeneous model and a log-uniform prior on the internal luminosity parameters, the homogeneous model and a linear-uniform prior on the internal luminosity parameters, and the inhomogeneous model and a linear-uniform prior on the internal luminosity parameters.}
    \label{tab:results_WASP103b}
\end{table*}




\bsp	
\label{lastpage}
\end{document}